\documentclass[a4paper,twocolumn,prb,showpacs,amssymb,aps,superscriptaddress]{revtex4}
\usepackage{dcolumn}
\usepackage{bm}
\usepackage{graphicx}
\def\tco {$T_{\rm CO}$}
\def\tc {$T_{\rm C}$}
\def\tjt {$T_{\rm JT}$}

\newcommand{\lsmo}{$\mathrm{La_{1-x}Sr_xMnO_{3}}$}
\newcommand{\lpsmo}{$\mathrm{(La_{1-y}Pr_y)_{7/8}Sr_{1/8}MnO_{3}}$}
\newcommand{\lpten}{$\mathrm{(La_{0.9}Pr_{0.1})_{7/8}Sr_{1/8}MnO_{3}}$}
\newcommand{\lseight}{$\mathrm{La_{7/8}Sr_{1/8}MnO_{3}}$}
\newcommand{\lsno}{$\mathrm{La_{1.67}Sr_{0.33}NiO_{4}}$}

\newcommand{\LPSMOa}{(La$_{0.9}$Pr$_{0.1}$)$_{7/8}$Sr$_{1/8}$MnO$_3$}

\newcommand{\LCMOx}{La$_{1-x}$Ca$_{x}$MnO$_3$}
\newcommand{\PCMOx}{Pr$_{1-x}$Ca$_{x}$MnO$_3$}
\newcommand{\LSMO}{La$_{\mbox{\scriptsize 7/8}}$Sr$_{\mbox{\scriptsize 1/8}}$MnO$_{\mbox{\scriptsize 3}}$}
\newcommand{\LPSMO}{(La$_{\mbox{\scriptsize 1-z}}$Pr$_{\mbox{\scriptsize z}}$)$_{\mbox{\scriptsize 7/8}}$Sr$_{\mbox{\scriptsize 1/8}}$MnO$_{\mbox{\scriptsize 3}}$}
\newcommand{\LSMOx}{La$_{\mbox{\scriptsize 1-x}}$Sr$_{\mbox{\scriptsize x}}$MnO$_{\mbox{\scriptsize 3}}$}

\newcommand{\figref}[1]{Fig.~\protect\ref{#1}}

\begin{document}

\title{Pressure-induced melting of the orbital polaron lattice in \LSMOx}
\author{R. Klingeler}\email[]{r.klingeler@ifw-dresden.de}
\affiliation{Leibniz-Institute for Solid State and Materials Research IFW Dresden, 01171 Dresden,
Germany}
\author{J. Geck}
\affiliation{Leibniz-Institute for Solid State and Materials Research IFW Dresden, 01171 Dresden,
Germany}
\author{S. Arumugam}\affiliation{Leibniz-Institute for Solid State and Materials Research IFW Dresden, 01171 Dresden,
Germany} \affiliation{High Pressure Low Temperature Lab, School of Physics, Bharathidasan
University, Tiruchirappalli - 620024, India}
\author{N. Tristan}
\affiliation{Leibniz-Institute for Solid State and Materials Research IFW Dresden, 01171 Dresden,
Germany}
\author{P. Reutler}\affiliation{Leibniz-Institute for Solid State and Materials Research IFW Dresden, 01171 Dresden,
Germany}\affiliation{Laboratoire de Physico-Chimie de l'Etat Solide, Universit\'e Paris-Sud XI,
91405 Orsay C\'edex, France}
\author{B. B\"{u}chner}
\affiliation{Leibniz-Institute for Solid State and Materials Research IFW Dresden, 01171 Dresden,
Germany}
\author{L. Pinsard-Gaudart}\author{A. Revcolevschi}
\affiliation{Laboratoire de Physico-Chimie de l'Etat Solide, Universit\'e Paris-Sud XI, 91405
Orsay C\'edex, France}

\date{\today}

\begin{abstract}
We report on the pressure effects on the orbital polaron lattice in the lightly doped manganites
\lsmo , with $x\sim 1/8$. The dependence of the orbital polaron lattice on $negative$ chemical
pressure is studied by substituting Pr for La in \LPSMO . In addition, we have studied its
hydrostatic pressure dependence in \lpten . Our results strongly indicate that the hopping $t$
significantly contributes to the stabilization of the orbital polaron lattice and that the orbital
polarons are ferromagnetic objects which get stabilized by local double exchange processes. The
analysis of short range orbital correlations and the verification of the Gr{\"u}neisen scaling by hard
x-ray, specific heat and thermal expansion data reinforces our conclusions.
\end{abstract}

 \maketitle

\section{INTRODUCTION}

The observation of a ferromagnetic insulating phase in the low-doped manganites \lsmo\ with $x\sim
1/8$ raised a large number of studies aimed to explain the properties of this phase (e.g.
\onlinecite{endoh99,nojiri99,uhlenbruck99,Ivanshin00,Deisenhofer05}). Several studies have shown
that charge and orbital ordering phenomena are crucial to understand the obvious contradiction to
the bare double exchange (DE) model, which predicts metallic behavior in case of ferromagnetic
spin order. Recently, resonant x-ray diffraction succeeded to detect the formation of an orbital
polaron lattice (OPL) at low temperatures, which unifies ferromagnetic and insulating properties
in a natural way. It was suggested that local charge hopping processes contribute significantly to
the stabilization of the orbital polarons. Further support for this conclusion was found in a
recent thermodynamic study which proved the magnetic DE energy to be crucial for the stabilization
of the OPL.\cite{klingeler02}

The substitution of La by smaller Pr, i.e. the application of chemical pressure, provides a direct
route to influence the balance between the various couplings that stabilize the OPL in the FMI
phase. More specifically, Pr doping causes an increase of the octahedral tilt angles, thereby
reducing the Mn-O-Mn bond angles and the hopping term $t$. It is well known that this structural
change dramatically alters the relative stability of different possible phases.
\cite{TokuraScience00} In particular, metallic phases with delocalized charges are suppressed upon
decreasing $t$.

In contrast to the application of negative chemical pressure, where different samples with
different amounts of Pr are studied, hydrostatic positive pressure effects can be studied on a
single particular compound. Here, we present data on the doping dependence of the ordering
phenomena in \LPSMO , with $y$=0, 0.1, 0.25, 0.5, 0.75, and on their hydrostatic pressure
dependence in \lpten . Our data imply that the OPL phase is stabilized by hydrostatic pressure,
i.e. upon increasing the charge mobility $t$. Whereas, application of negative chemical pressure
yields the opposite results. The verification of the Gr{\"u}neisen scaling and the analysis of short
range orbital correlations reinforces our conclusions. Our study hence clearly confirms the
microscopic OPL model. We mention that recently ferromagnetic correlations have been found in the
charge stripe ordered phase of \lsno , which might be due to local DE, too.\cite{klingelerni} The
application of chemical or external pressure might hence also affect the ferromagnetism in the
doped nickelates as it is well known in the low-doped manganites.

\section{Experimental and results}

\subsection{Experimental}

We report on measurements of the magnetisation, the specific heat, the thermal expansion, and on
high energy x-ray diffraction data of \lpsmo\ single crystals in external magnetic fields up to
16\,T. The experiments were carried out on single crystals, with $y$=0, 0.1, 0.25, 0.5, 0.75,
grown by the floating zone method.\cite{Pascal} The magnetisation $M(B)$ was measured with a SQUID
and a vibrating sample magnetometer. Magnetization measurements under hydrostatic pressure were
performed in the temperature range 4.2 - 350K using an MPMS SQUID magnetometer (Quantum Design)
and a miniature piston-cylinder pressure cell made of Cu-Be alloy (PSM 100, Tetcon International).
Silicon oil was used as the pressure transmitting medium. Constant magnetic fields of 0.05\,T in
the temperature region 50 - 250K and of 0.5\,T were used in the temperature range 250 - 350\,K for
all measurements under pressure. The pressure at low temperature was determined by the pressure
dependence of superconducting transition temperature of high purity Pb placed with the sample.

The specific heat $c_p$ was measured using a high resolution calorimeter. Here, we have applied
two different quasi-adiabatic methods, continuous heating and heating pulses.~\cite{klingeler02}
For the thermal expansion studies we applied a high resolution capacitive
dilatometer.\cite{lorenz97} Hard x-ray data have been measured at the beamline BW5 at the HASYLAB,
Hamburg. For experimental details see Ref.~\onlinecite{geckNJP04}.

\subsection{Pressure dependence of orbital order in \lpsmo }

\begin{figure}[t!]
\center{\includegraphics*[angle=-90,width=0.95\columnwidth]{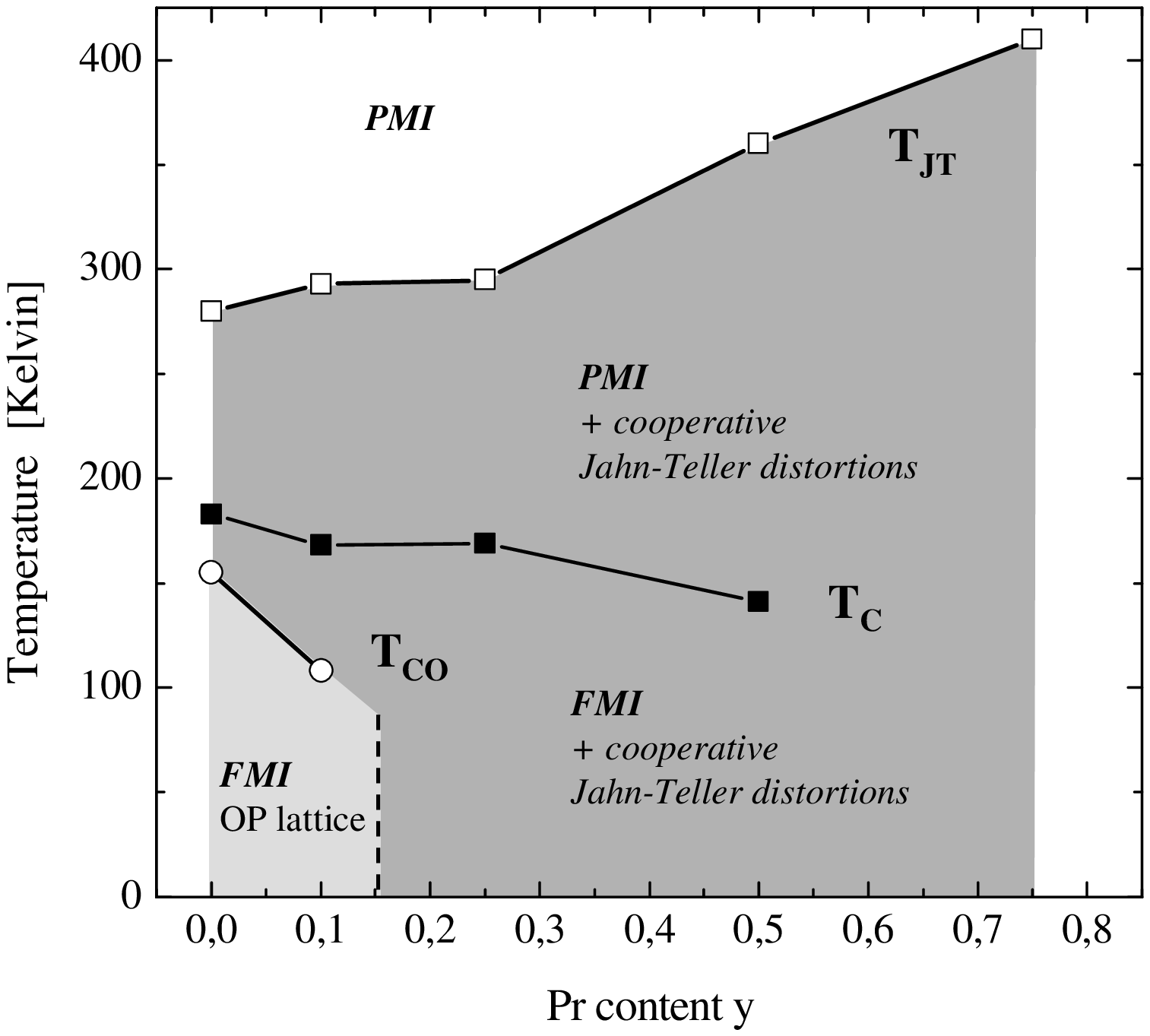}}
  \caption{\label{pd}Phase diagram of \LPSMO, based on x-ray powder diffraction as well as magnetization and resistivity measurements.
The suppression of the orbital polaron lattice upon Pr doping, i.e. absence of the corresponding
superstructure reflections, has been verified by high energy x-ray diffraction. \tjt\ and \tco\
denote the transition temperatures between the indicated phases. (PMI: paramagnetic insulating;
FMI: ferromagnetic insulating; FMM: ferromagnetic metallic)}
\end{figure}

The phase diagram of \LPSMO\ as it is presented in \figref{pd} was established by measurements of
the magnetization, the specific heat, the electrical resistivity, and by x-ray diffraction data.
For $y=0$, several ordering phenomena are observed upon cooling. At \tjt\ = 270\,K , a cooperative
Jahn-Teller effect evolves, and ferromagnetic spin occurs at \tc\ = 183\,K. Eventually, the
ferromagnetic insulating phase exhibiting the orbital polaron lattice is realized below \tco\ =
150\,K. Upon Pr doping, i.e. by applying chemical pressure, all ordering temperatures
significantly change. (1) The orbital polaron lattice is rapidly suppressed upon Pr doping.
Already for $y=0.1$ the orbital polaron lattice is considerably destabilized, while it is
completely suppressed for $y=0.25$. (2) \tc\ is suppressed, too, but the effect on the
ferromagnetic spin ordering is less than on \tco . (3) \tjt\ increases upon Pr doping, i.e. the
cooperative Jahn-Teller distortions are stabilized.

\begin{figure}[t]
\center{\includegraphics*[width=0.95\columnwidth]{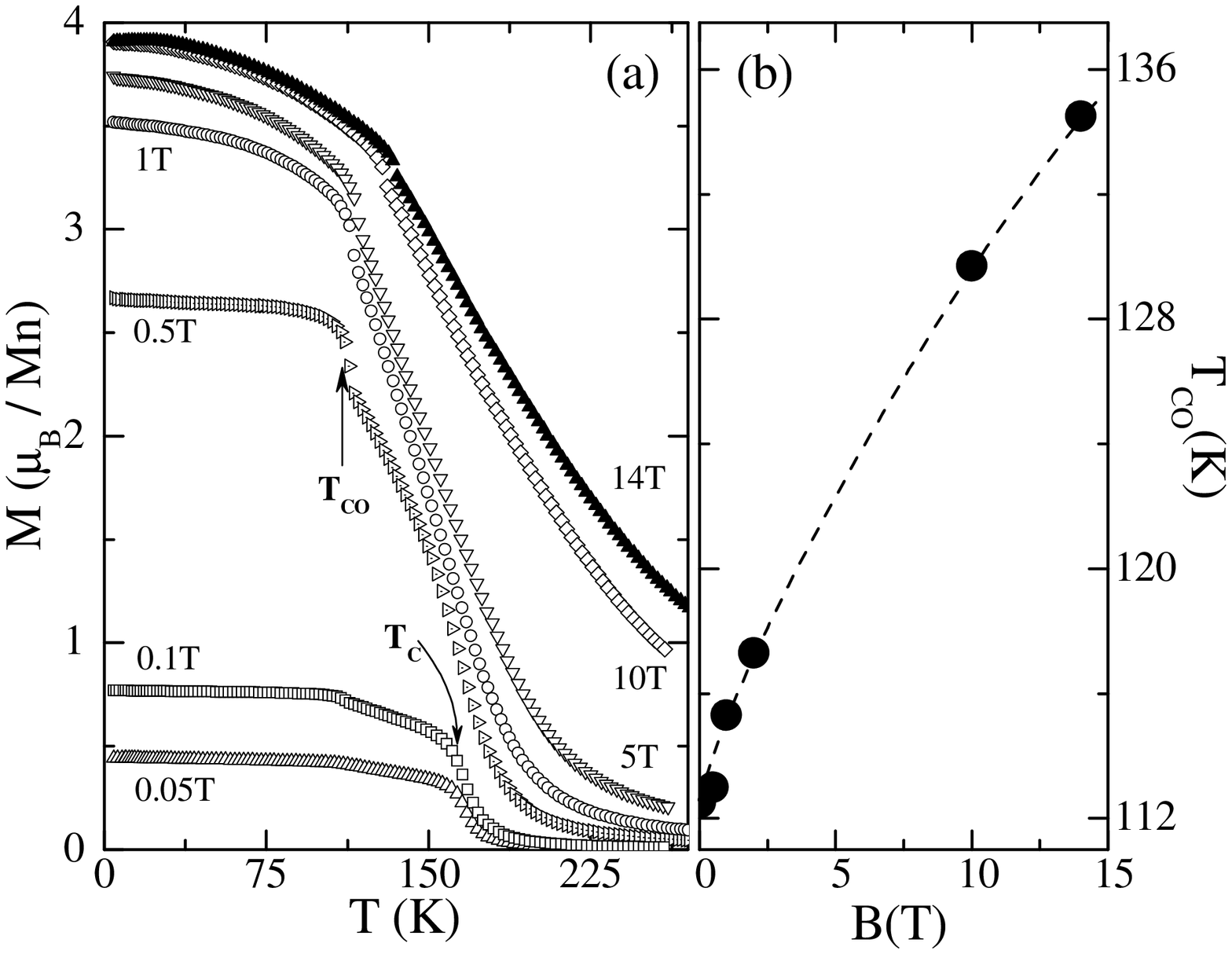}}
  \caption{(a) Magnetization vs. temperature, at different magnetic fields up to 14\,T, of \lpten . \tc\ and \tco\ label the onset
  of ferromagnetic spin ordering and the OPL phase at \tco . (b) Field dependence of \tco\ as extracted from the magnetization data.}
\label{mag}
\end{figure}

In \figref{mag}(a) we show the effect of the spin and the orbital ordering on the magnetization of
\lpten , under various applied magnetic fields up to 14\,T. The effect of the magnetic field on
the OPL phase is also illustrated by \figref{mag}(b) which displays the field dependence of \tco .
At low temperatures the ferromagnetic OPL phase is realized. Upon heating, the orbital polaron
lattice melts at \tco\ $\simeq$ 110\,K, at which the transition is associated with a kink in the
magnetization $M(T)$. This kink is visible in \figref{mag}(a) e.g. at $B=0.5$\,T. The measurements
in different magnetic fields imply that \tco\ increases by 22\,K upon application of $B=14$\,T
[see \figref{mag}(b)]. A similar behavior was observed for \lseight\ (see Fig.\,6 of
Ref.~\onlinecite{klingeler02}). When the temperature is further increased, the ferromagnetic spin
order vanishes at \tc\ $\simeq$ 168\,K, and the structural Jahn-Teller phase transition is
observed at \tjt\ $\simeq$ 296\,K.\cite{geckNJP04}

\begin{figure}[t]
\center{\includegraphics*[width=0.9\columnwidth]{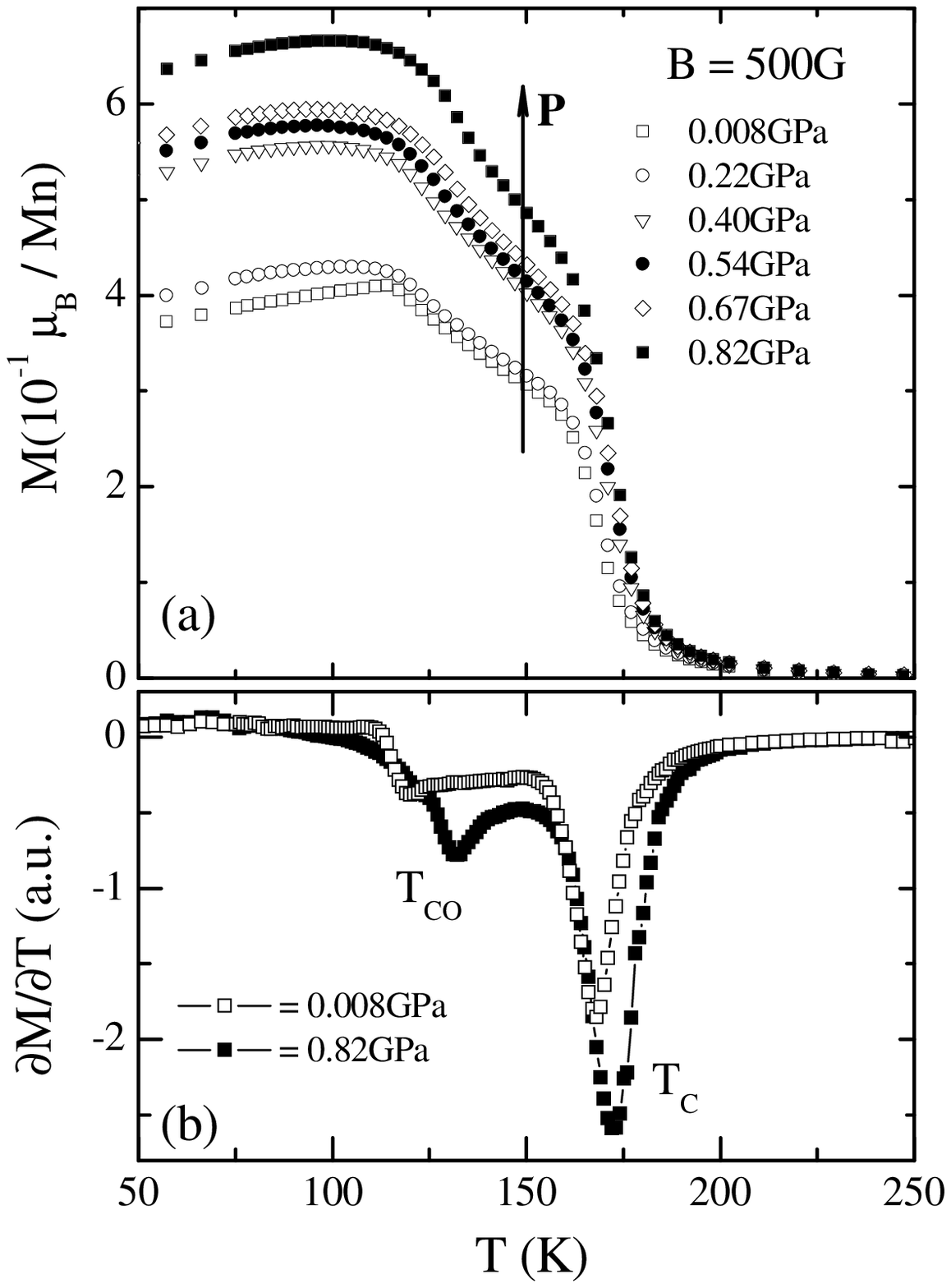}}
  \caption{(a) Magnetization vs. temperature at different external hydrostatic pressure ($B=500$\,G).
  (b) Derivative $\partial M/\partial T$ for lowest and highest applied pressure.}
\label{pressure1}
\end{figure}

\begin{figure}[t]
\center{\includegraphics*[width=0.9\columnwidth]{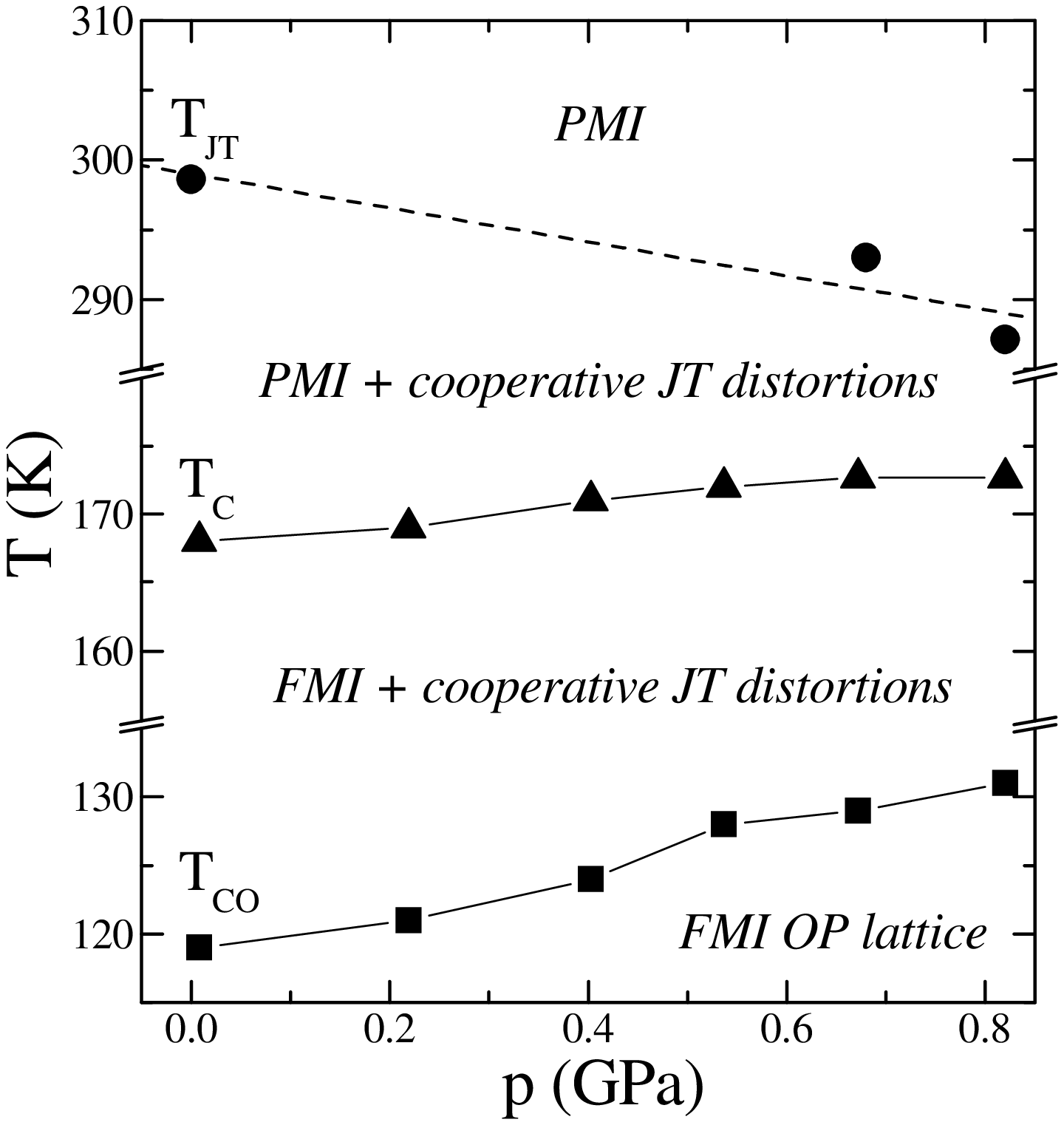}}
  \caption{Hydrostatic pressure dependencies of \tjt , \tc\ and \tco , as extracted from the magnetization data in \figref{pressure1}.}
\label{pressure2}
\end{figure}

In order to study the effect of (\textit{positive}) hydrostatic pressure on the orbital and spin
ordering phenomena, we studied the temperature dependence of the magnetization of \LPSMOa\ single
crystals at various hydrostatic pressures up to 0.82\,GPa [see \figref{pressure1}(a)]. The data
show the increase of the magnetization in the ferromagnetic metallic (FMM) and in the
ferromagnetic OPL phase upon application of external hydrostatic pressure, whereas there are only
minor pressure effects on the paramagnetic phase at $T>T_{\rm C}$. From the data in
\figref{pressure1}(a), we have derived the pressure dependencies of the ordering temperatures \tco
, \tc\ and \tjt , respectively, which are displayed in \figref{pressure2}. The data show that
\tco\ and \tc\ increase upon applying hydrostatic pressure, whereas \tjt\ becomes smaller.
Thermodynamically, this behavior is in accordance with the fact that the volume of the unit cell
shrinks at the Jahn-Teller transition and increases at \tc\ and \tco .\cite{geckNJP04}

The fact that both the magnetization below \tc\ down to 50\,K and \tc\ itself increase as a
function of pressure is well understood in the framework of the DE model. Enhancement of \tc\
under hydrostatic pressure is a common trend of perovskite manganites, whereas in the single
layered manganites uniaxial pressure is necessary to affect the spin ordering phenomena
significantly.\cite{Zhou,klingelerJEMS} In the case of \LPSMOa , the hydrostatic pressure yields
an increase of $t$, thereby stabilizing the FMM ordered phase by enhancing the mobility of the
charge carriers via the DE. Remarkably, the hydrostatic pressure dependence not only of \tc\ but
also of \tco\ is positive. This observation proves the important role of the charge hopping $t$
and the DE for the stabilization of the OPL. The comparison of the magnetic field and the
hydrostatic pressure effect on \tco\ suggests that the application of $B=5.6$\,T is approximately
equivalent to the application of 1\,GPa pressure.

At ambient pressure, the structural phase transition to the co-operative Jahn-Teller-distorted
phase occurs close to room temperature, i.e. \tjt\ $\simeq$ 296\,K, for $y=0.1$. Under applied
hydrostatic pressure, i.e. if the hopping parameter $t$ and hence the charge mobility is
increased, the cooperative JT distorted phase becomes suppressed and the dynamic JT phase is
stabilized. These effects are just opposite to the chemical pressure effect upon Pr doping
($y$=0.1) at the La site. Note, that the effect of destabilization of the cooperative JT phase and
stabilization of dynamic JT phase is observed in \lsmo\ under applied magnetic field,
too.\cite{uhlenbruck99}

A striking feature of our data is that, at high pressure, \tco\ increases faster than \tc\ under
pressure. This means that changes of $t$ influence the OPL phase more effectively than the FMM
phase, i.e. the DE promotes the OPL even more efficiently than the FMM phase. Faster increase of
\tco\ than \tc\ was reported under pressure in the low-doped regime around $x=1/8$.\cite{Zhou} For
small external pressure, however, the effect on \tco\ is similar to that on \tc , which agrees
with the Gr{\"u}neisen scaling presented in Sec.\,\ref{Secgruen}.

\subsection{Doping dependence of the OPL phase}

The experimental data presented in the previous sections show that chemical (i.e.
\textit{negative} hydrostatic) pressure due to the substitution of Pr causes the destabilization
of the OPL phase in \lseight , whereas  \textit{positive} hydrostatic pressure yields the opposite
effect. We have shown experimentally that the hopping $t$ significantly contributes to the
stabilization of the OPL. This fact provides a natural explanation for the absence of the OPL in
the FMI-phase of lightly doped \LCMOx\ and \PCMOx , where $t$ is reduced due to the smaller Pr and
Ca sites. For these systems there is evidence that a FMI and cooperative Jahn-Teller distorted
phase exists, which corresponds to the low temperature phase of \LPSMO\ with $y\geq 0.25$.

\begin{figure}[t!]
\center{\includegraphics*[angle=-90,width=0.99\columnwidth]{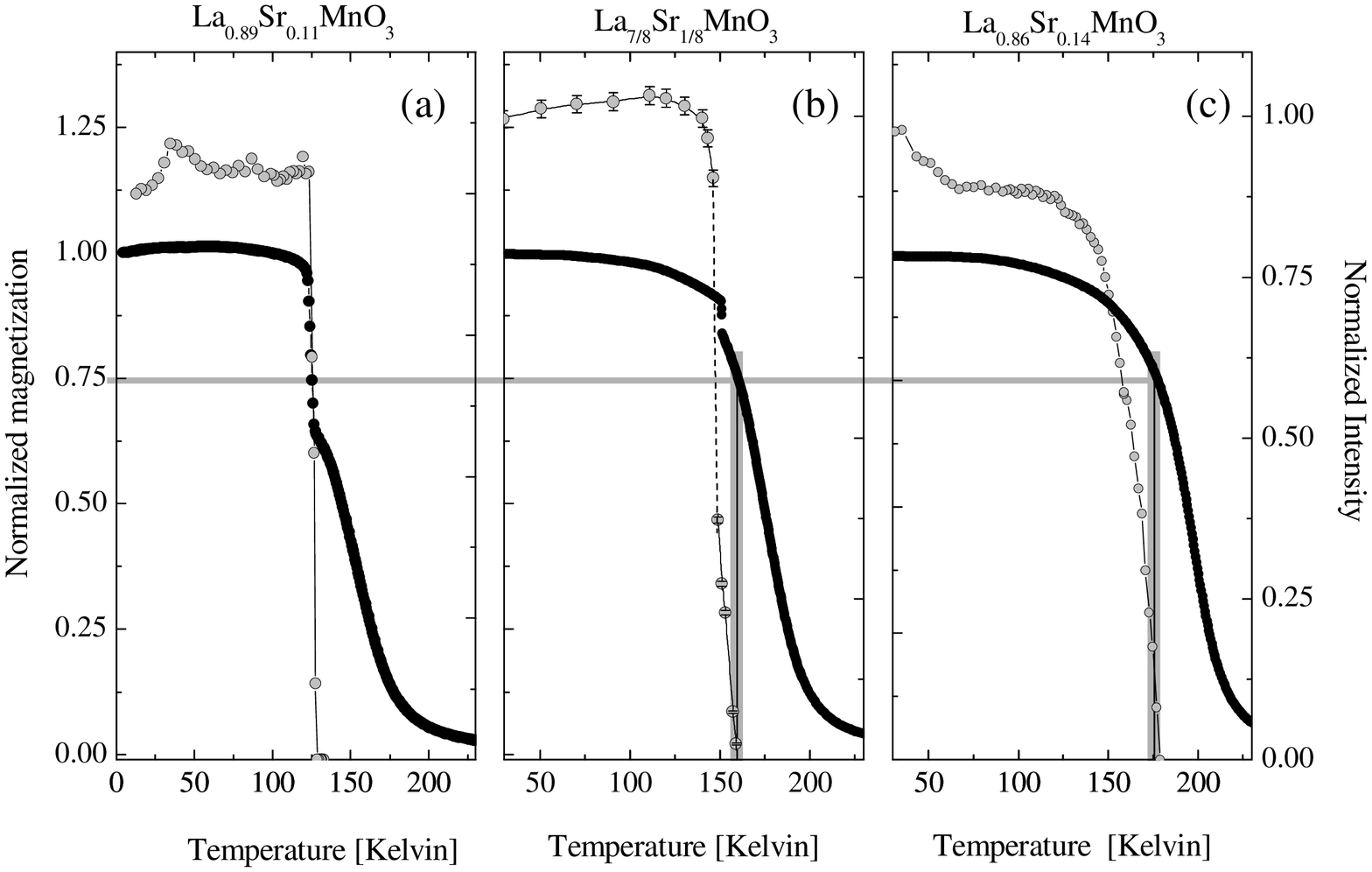}}
  \caption{Comparison between the temperature dependencies of the
    magnetization measured in an external field of 0.5\,Tesla and the intensities of the superlattice reflections
    determined by HXS for $x=0.11$, 1/8 and 0.14. The data sets have been
    normalized to the respective low temperature values. The superlattice
    reflections only occur, if the normalized magnetization increases
    above 0.75 (gray lines).}
\label{fig:dot_vgl_hxs_m}
\end{figure}

The relevance of the charge hopping $t$ for the stabilization of the OPL  in \LSMOx\ is further
supported by the measurements shown in Fig.\,\ref{fig:dot_vgl_hxs_m}. In this figure, the
temperature dependencies of the magnetization measured in an external field of 0.5\,Tesla and the
intensities of the superlattice reflections measured by high-energy x-ray diffraction are compared
for $x=0.11$, 1/8 and 0.14.
%
%
The magnetic entropy in the FMM-phase between \tc\ and \tco\ decreases with increasing doping
level; i.e. the magnetic disorder in this phase is largest for $x=0.11$ and decreases towards
$x=0.14$. According to the double exchange (DE), the charge hopping in the FMM-phase of
La$_{0.89}$Sr$_{0.11}$MnO$_3$ is considerably suppressed, as the magnetic disorder is large. At
the same time, the intensity of the superstructure reflection corresponding to the OPL vanishes
completely in the FMM-phase [Fig.\,\ref{fig:dot_vgl_hxs_m}(a)] of this compound.
%
The increase of the strontium doping to $x=1/8$ leads to a reduction of the magnetic disorder in
the FMM-phase as compared to the $x=0.11$ sample. In this case, the superstructure reflection
corresponding to the OPL is already observed in the FMM-phase [Fig.\,\ref{fig:dot_vgl_hxs_m}(b)],
revealing the presence of short range OPL order in the FMM-phase of \LSMO. Similarly, the
HXS-measurements on La$_{0.86}$Sr$_{0.14}$MnO$_3$ indicate the presence of short range OPL
correlations  above \tco , too [cf. Fig.\,\ref{fig:dot_vgl_hxs_m}(c)].

A remarkable observation concerns the temperatures which mark the onset of the short range
correlations. As indicated in Fig.\,\ref{fig:dot_vgl_hxs_m}, the onset of the short range
correlations in the samples with $x=1/8$ and $x=0.14$ corresponds in both cases to the temperature
where the magnetization reaches about 75\,\% of its saturation value. For $x=0.11$,  the
magnetization never reaches the 75\,\% value (gray horizontal line in
Fig.\,\ref{fig:dot_vgl_hxs_m}) in the FMM phase of $x=0.11$; i.e. the magnetic disorder is too
large to allow for the short range OPL order.

%

\subsection{\label{Secgruen}Gr{\"u}neisen scaling for \lseight }

\begin{figure}[t]
\center{\includegraphics [width=0.9\columnwidth,clip] {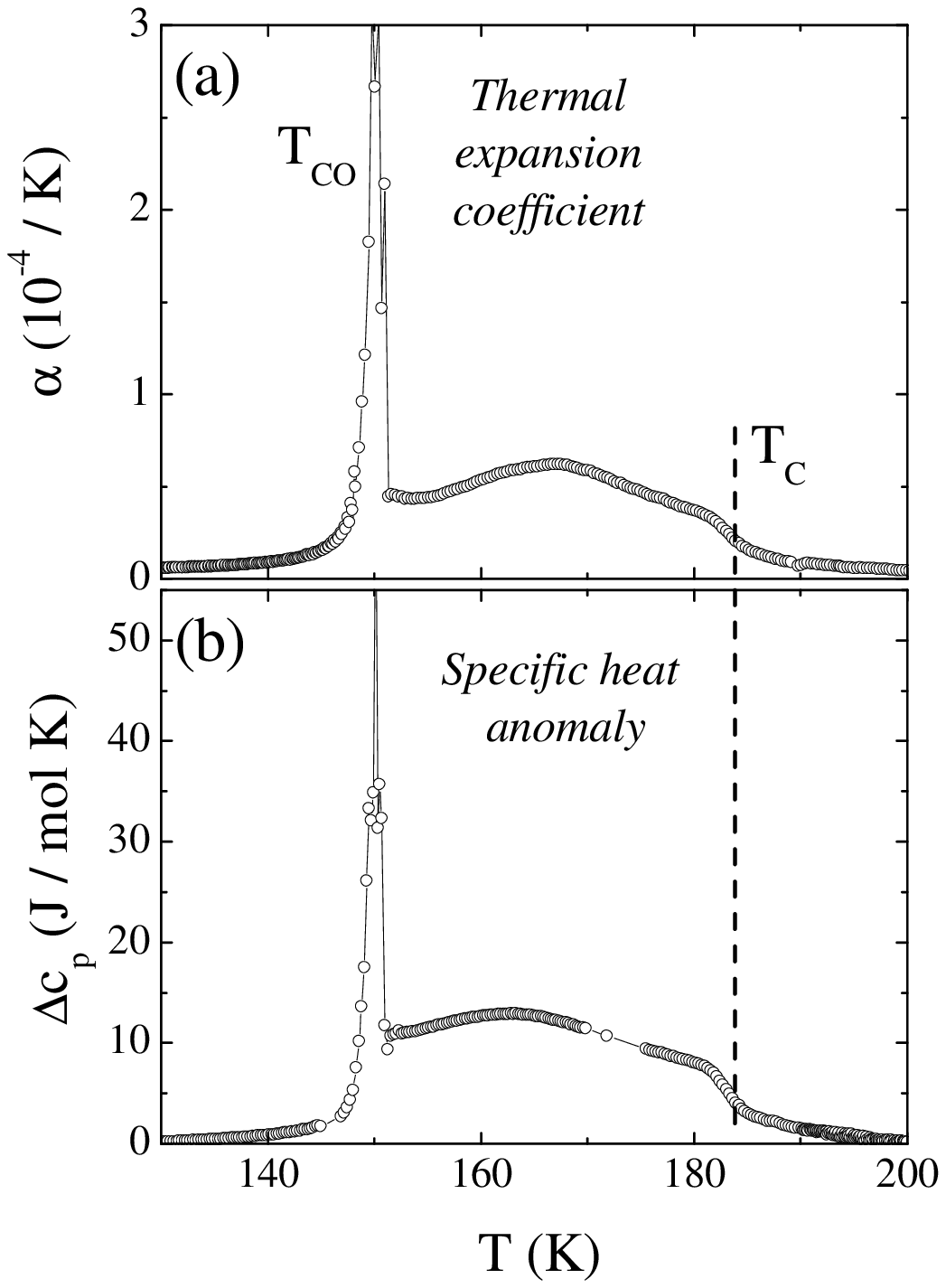}} \caption[]
{\label{gruen} (a) Thermal expansion coefficient and (b) specific heat anomaly of \lseight . }
\end{figure}

The results presented above clearly show that the formation of the OPL phase and of the FMM phase
is based on a common energy scale which is given by the hopping parameter $t$. Hence, the pressure
dependencies of the relevant energy scale for the ordering phenomena must agree, since it is the
same energy scale. This can be checked experimentally by the verification of the Gr{\"u}neisen
scaling, i.e by measuring the anomalous contributions to the specific heat $c_p$ and to the
thermal expansion coefficient $\alpha$. If the Gr{\"u}neisen scaling is fulfilled, both quantities are
proportional. Quantitatively, the Gr{\"u}neisen relation reads

\begin{equation}
\frac{\alpha}{c_p}=\frac{1}{V}\left. \frac{\partial \ln \epsilon}{\partial p}\right|_T,
\end{equation}

with $\epsilon$ being the relevant energy scale of the respective ordering phenomenon, i.e. in the
case at hand $\epsilon=\epsilon(t)$. We hence studied the specific heat and the thermal expansion
coefficient of \lseight . From both data we subtracted the phonic contribution, which was
estimated from the specific heat data as is explained e.g. in Ref. \onlinecite{klingeler02}. The
resulting anomalous contributions to the specific heat and to the thermal expansion are shown in
\figref{gruen}. As has been shown earlier\cite{klingeler02,uhlenbruck99}, the formation of the FMM
phase causes a jump of the specific heat at \tc\ and a large peak is associated with the onset of
the OPL phase at \tco . In addition, there is a regime of anomalous entropy changes at \tco\ $\leq
T \leq$ \tc . All these features are visible in the thermal expansion data in \figref{gruen}(a),
too. There is a peak at \tco , a jump at \tc\ and a regime of anomalous length changes in the
whole FMM phase. The data hence clearly prove very similar temperature dependencies of $c_p$ and
$\alpha$. This observation implies that both ordering phenomena at \tco\ and at \tc\ have the same
pressure dependence, which reinforces our conclusions drawn above. In addition, this is also true
for the entropy changes in the whole temperature regime \tco\ $\leq T \leq$ \tc .

\begin{figure}
\center{\includegraphics [width=1.0\columnwidth,clip] {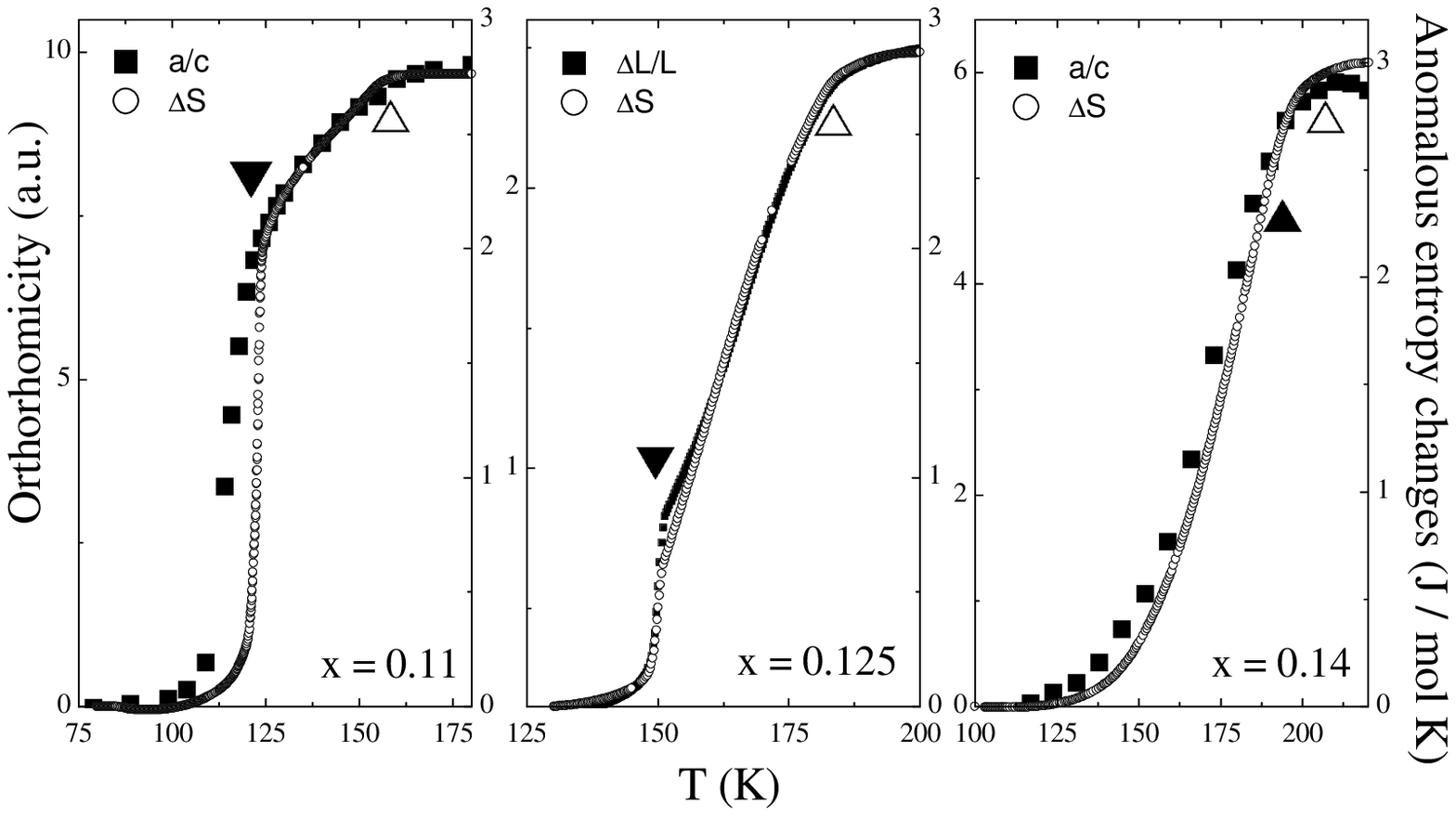}} \caption[]
{\label{gruen2} Temperature dependence of the orthorhombicity $a/c$ and of the anomalous entropy
changes for \lsmo , with $x=0.11$ and $x=0.14$. For comparison, the integrated data of
\figref{gruen}, i.e. the anomalous length and entropy changes for $x=1/8$, are also shown. Full
(open) triangles mark \tco\ (\tc ).}
\end{figure}

This conclusion does not only hold for \lseight\ but also for $x=0.11$ and 0.14. This is
illustrated by \figref{gruen2}, which shows that the Gr{\"u}neisen scaling is valid in its integral
form for these compositions. The figure presents the temperature dependencies of the
orthorhombicity $a/c$ and of the anomalous entropy changes $\Delta S = \int c_p/T dT$.

\section{Conclusion}

Our results show that charge hopping processes are essential to stabilize the OPL in \LSMOx.
However, since the OPL leads to macroscopically insulating behavior, these hopping processes take
place on a local scale, i.e. within the orbital polarons. To be more specific, the orbital
polarons are ferromagnetic objects which get stabilized by local DE processes.\cite{Kilian99} The
pressure effects enhance the local DE processes by increasing $t$ and hence lead to the
stabilization of OPL at low temperature. The magnetic disorder and the suppression of the orbital
polaron lattice created by the Pr ($y$=0.1) doping is counterbalanced by the external pressure for
stabilizing the orbital polaron lattice at low temperatures. Our analysis of the short range
orbital correlations and the verification of the Gr{\"u}neisen scaling reinforces our conclusions
regarding the relevance of $t$ and DE for the formation of the OPL.\\

\begin{acknowledgments}
This work was partially supported by DFG-Indian National Science Academy International Exchange
fellowship program.
\end{acknowledgments}

\end{document}